\documentclass[preprint,12pt]{elsarticle}

\usepackage{amssymb}

\journal{Synthetic Metals}

\begin{document}
\newcommand{\icm}{\ensuremath{\mbox{cm}^{-1}}}
\newcommand{\SF}{$\rm (TM\-TSF)_2\-PF_6$}
\newcommand{\TF}{$\rm (TM\-TTF)_2\-PF_6$}
\newcommand{\asf}{$\rm (TM\-TTF)_2\-AsF_6$}
\begin{frontmatter}



\title{Pressure-induced structural phase transition in the Bechgaard-Fabre salts}


\author[Augsburg]{A. Pashkin}
\author[Stuttgart]{M. Dressel}
\author[Halle]{S. G. Ebbinghaus}
\author[Grenoble]{M. Hanfland}
\author[Augsburg]{C. A. Kuntscher}

\address[Augsburg]{Experimentalphysik II, Universit\"at Augsburg,
Universit\"atsstr. 1, 86159 Augsburg, Germany}

\address[Stuttgart]{1. Physikalisches Institut, Universit\"at Stuttgart,
Pfaffenwaldring 57, 70550 Stuttgart, Germany}

\address[Halle]{Martin-Luther-Universit\"at Halle-Wittenberg, Institut f\"ur
Chemie, Kurt-Mothes-Stra{\ss}e 2, 06120 Halle, Germany}

\address[Grenoble] {European Synchrotron Radiation Facility, BP 220, 38043
Grenoble, France}

\begin{abstract}
The crystal structures of the quasi-one-dimensional organic salts
(TMTTF)$_2$\-PF$_6$ and (TMTSF)$_2$\-PF$_6$ were studied by
pressure-dependent x-ray diffraction up to 10 GPa at room
temperature. The unit-cell parameters exhibit a clear anomaly due to
a structural phase transition at 8.5 and 5.5~GPa for
(TMTTF)$_2$\-PF$_6$ and (TMTSF)$_2$\-PF$_6$, respectively.

\end{abstract}


\end{frontmatter}


\section{Introduction}

The Fabre salts (TMTTF)$_2X$ and the Bechgaard salts (TMTSF)$_2X$
represent prime examples of quasi-one-dimensional electronic
systems demonstrating a large variety of phenomena such as
superconductivity, charge order, Mott-Hubbard insulating state,
spin-density-waves etc. which have been extensively studied over the
last three decades \cite{Jerome82,Ishiguro98,Jerome04,Dressel03}.
The structure of these salts consist of molecular stacks formed by
tetra\-methyl\-tetra\-thia\-fulvalene (TMTTF) or
tetra\-methyl\-tetra\-selena\-fulvalene (TMTSF) cations separated by
the monovalent $X^-$ anions. At room temperature the (TMTTF)$_2X$ salts are
typically Mott-Hubbard insulators due to the strong correlation of
electrons and a weak interchain coupling \cite{Giamarchi07}. On the
other hand, the (TMTSF)$_2X$ salts are good quasi-one-dimensional
metals although their properties are still strongly affected by
correlation effects \cite{Dressel03,Dressel96}.

In the Bechgaard-Fabre salts
the interstack separation and therefore the interchain hopping
integral can be nicely tuned either by
changing the size of the $X^-$ anions or the type of the cation (chemical
pressure effect), or by applying external pressure. It has
been demonstrated that the effect of chemical and external
hydrostatic pressure are virtually equivalent; i.e.\
numerous physical quantities of (TMTSF)$_2X$ resemble those of (TMTTF)$_2X$
under pressure. This provides the possibility to
construct a generic temperature-pressure phase
diagram \cite{Bourbonnais98,Wilhelm01} of the Bechgaard-Fabre salts.
Continuous tuning of external pressure and temperature is the best
way to explore the phase boundaries in this diagram.
Jaccard et al. \cite{Jaccard01} employed a diamond anvil cell (DAC) to generate extreme
pressures that made it possible, for instance, to tune \TF\ throughout the
phase diagram starting from the insulating spin-Peierls up to the
superconducting state. A DAC has also been used in
infrared spectroscopy experiments that demonstrated
pressure-induced insulator-to-metal transition in the TMTTF
family \cite{Pashkin06,Pashkin09}. For a correct quantitative analysis
of data obtained in the above mentioned experiments, knowledge of the
crystal structure is crucial. However, the compressibility and
structure at high pressures of the Bechgaard-Fabre salts has not
been measured up to the present day. To the best of our knowledge, the highest
pressure at which the structure has been determined
is 1.6 GPa for the example of \SF\ - the first organic
superconductor \cite{Gallois86,Gallois87}. On the other hand,
high-pressure dc transport of \TF\ has been investigated at
pressures up to 8 GPa \cite{Jaccard01}, much higher than any
structural study of the Bechgaard-Fabre salts reported up to now.

In this paper, we present results of single-crystal x-ray
diffraction performed on two typical salts \TF\ and \SF\ for pressures
up to 10 GPa. We
determine the pressure dependence of the unit cell constants and
volume compressibility of the crystals. Moreover, we found
pressure-induced structural phase transitions which take place at
high pressures in both studied salts.

\begin{figure}[t]
  \centerline{
  \includegraphics[angle=270,width=0.75\columnwidth]{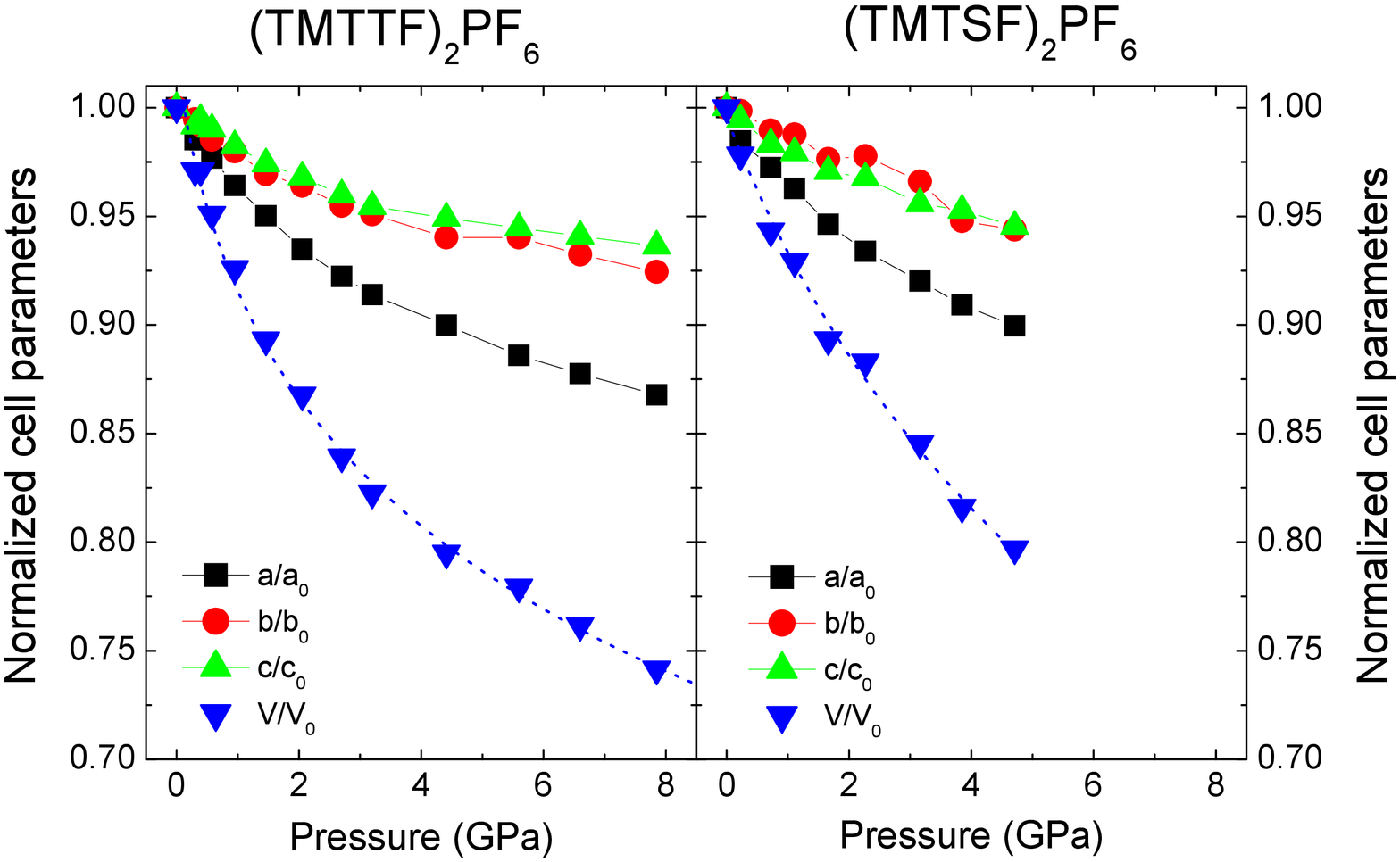}
  }\caption{(Color online) Normalized unit cell parameters of \TF\ and \SF\ as a function of pressure obtained at room temperature.
  Full lines are guides to the eye.
  The dotted blue lines correspond to the fit of the pressure-dependent
  unit-cell volume using the Birch equation~(\ref{eq:Birch}).
  }\label{fig:unitcell}
\end{figure}

\section{Experiment}

Single crystals of \TF\ and \SF\ were grown by a standard
electrochemical procedure \cite{Montgomery94}. The room-temperature
x-ray diffraction experiments were carried out at beamline ID09A of
the European Synchrotron Radiation Facility in Grenoble. The
wavelength used for the experiments was 0.413 $\rm \AA$. X-ray
diffraction patterns were collected on an image plate MAR345
detector. The DAC rotation angle varied from $-30^{\circ}$ to
$+30^{\circ}$ for \TF\ and from $-20^{\circ}$ to $+20^{\circ}$ for
\SF\ with $2^{\circ}$ step. Liquid helium served as pressure
transmitting medium in the DAC. The diffraction patterns have been
analyzed using the XDS package \cite{Kabsch93}.
The pressure in the DAC was determined {\it in situ} by the ruby
luminescence method \cite{Mao86}.

\section{Results and Discussion}

The single-crystal x-ray diffraction analysis confirms the
space-group symmetry $P\bar{1}$ of both studied compounds in the
low-pressure phase reported previously \cite{Delhaes79,Gallois86}.
The pressure dependence of the normalized cell parameters of
(TMTTF)$_2$PF$_6$ and (TMTSF)$_2$PF$_6$ compounds is shown in
Fig.~\ref{fig:unitcell}. The compression of both compounds is
anisotropic: The \textit{intrastack} molecular separation defined
by the unit cell parameter $a$ suffers an approximately two times
larger variation under pressure compared to the $b$ and $c$-axes
lattice parameters, which are related to the \textit{interstack}
separation. The maximal softness in the stacking direction is in
accord with the anisotropy of the thermal expansion coefficient
that is largest along the $a$ axis \cite{Gallois86,deSouza08}.

The change of the unit cell volume can be well fitted with the Birch
equation of state \cite{Birch78}:
\begin{equation}
  P(V)=\frac{3}{2}B_0(x^7-x^5)\left[1+\frac{3}{4}(B'_0-4)(x^2-1)\right]
  \label{eq:Birch}
\end{equation}
with $x=(V_0/V)^{1/3}$, where $V_0$ is the unit cell volume at
ambient pressure; $B_0$ denotes the bulk modulus and $B'_0$ its
pressure derivative. The parameters of the fit together with the
absolute values of the unit-cell dimensions are given in
Tab.~\ref{tab:Birch}. The values reported for \SF\ are in reasonable
agreement with previous structural studies under moderate
pressure \cite{Gallois86,Gallois87}.

\begin{table}[h]
\caption{\label{tab:Birch} The unit cell parameters of \TF\ and
\SF\ at ambient pressure, and the bulk modulus $B_0$ and its
pressure derivative $B'_0$ obtained from the fit according to
Eq.~(\ref{eq:Birch}).}
\begin{tabular}{ccc}
  & \TF  & \SF \\
\hline

$a$ (\AA) & 7.156 & 7.281 \\

$b$ (\AA) & 7.572 & 7.681 \\

$c$ (\AA) & 13.211 & 13.486 \\

$\alpha$ & 82.427 & 83.08 \\

$\beta$ & 84.661 & 86.322 \\

$\gamma$ & 72.321 & 70.829 \\

$V_0$ (\AA$^3$) & 675.02 & 706.95 \\

$B_0$ (GPa) & $7.27\pm 0.64$ & $12.71\pm 0.97$ \\

$B'_0$ & $9.98\pm 1.15$ & $4.23\pm 0.75$ \\

\end{tabular}
\end{table}

\begin{figure}[t]
  \centerline{
  \includegraphics[width=0.9\columnwidth]{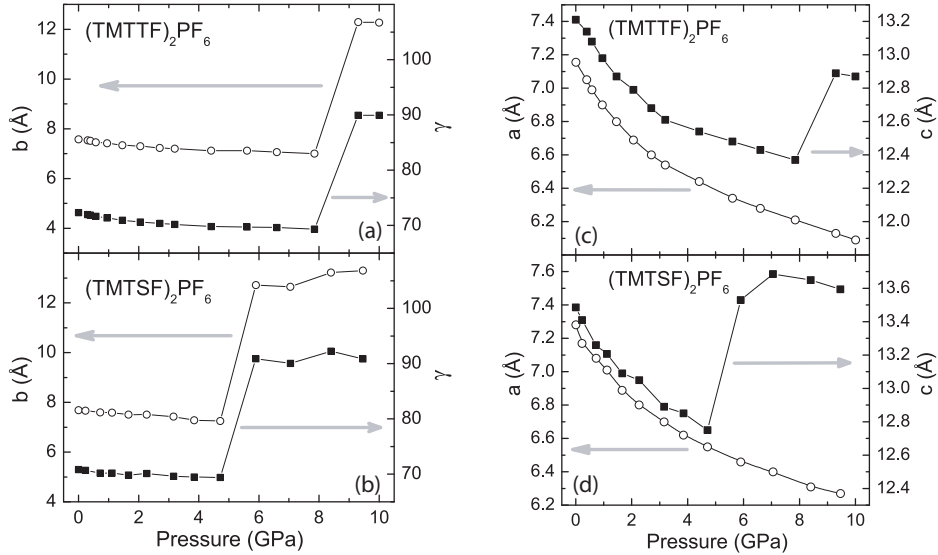}}
  \caption{Unit cell parameters of \TF\ [(a) and (c)] and \SF\ [(b) and (d)] as a function of
  pressure up to 10~GPa. Both the interstack separation $b$ [left axes of (a) and
  (b)]
  and the $\gamma$ angle [right axes of (a) and (b)] exhibit abrupt jumps that indicate
  structural phase transitions.
  The unit cell length along the longest interstack separation $c$ axis [right axes (c) and
  (d)]  shows an anomaly around the phase transition in both salts.
  At the same time the $a$ axis parameter does not reveal any noticeable discontinuity [left axes of (c) and
  (d)].}\label{fig:doubling}\label{fig:ac-transition}
\end{figure}

Upon increasing the pressure further, the unit-cell parameters
experience an abrupt jump indicating a pressure-induced structural
phase transition. The transition occurs at 5.5~GPa in \SF\ and at
8.5~GPa in \TF. The most salient changes across the transition are
the doubling of the unit cell along the $b$ axis and the increase
of the angle $\gamma$ to 90$^\circ$ (see Fig.~\ref{fig:doubling}),
which basically modifies the crystal symmetry from triclinic to
orthorhombic. This seems to be a stable configuration that does
not change with pressure any more.

The phase transition also affects the unit cell length along the $c$
axis, however, the $a$ axis unit cell length remains
unaffected. This is illustrated in Fig.~\ref{fig:ac-transition} for
both studied compounds. The abrupt increase of the $c$ axis dimension at
the transition pressure is of approximately 5\% and it is clearly
noticeable. The variation of the $a$ parameter seems to be smooth
across the whole studied pressure range without signs of any sharp
anomaly.

\begin{figure}[t]
  \centerline{
  \includegraphics[angle=270,width=0.8
  \textwidth]{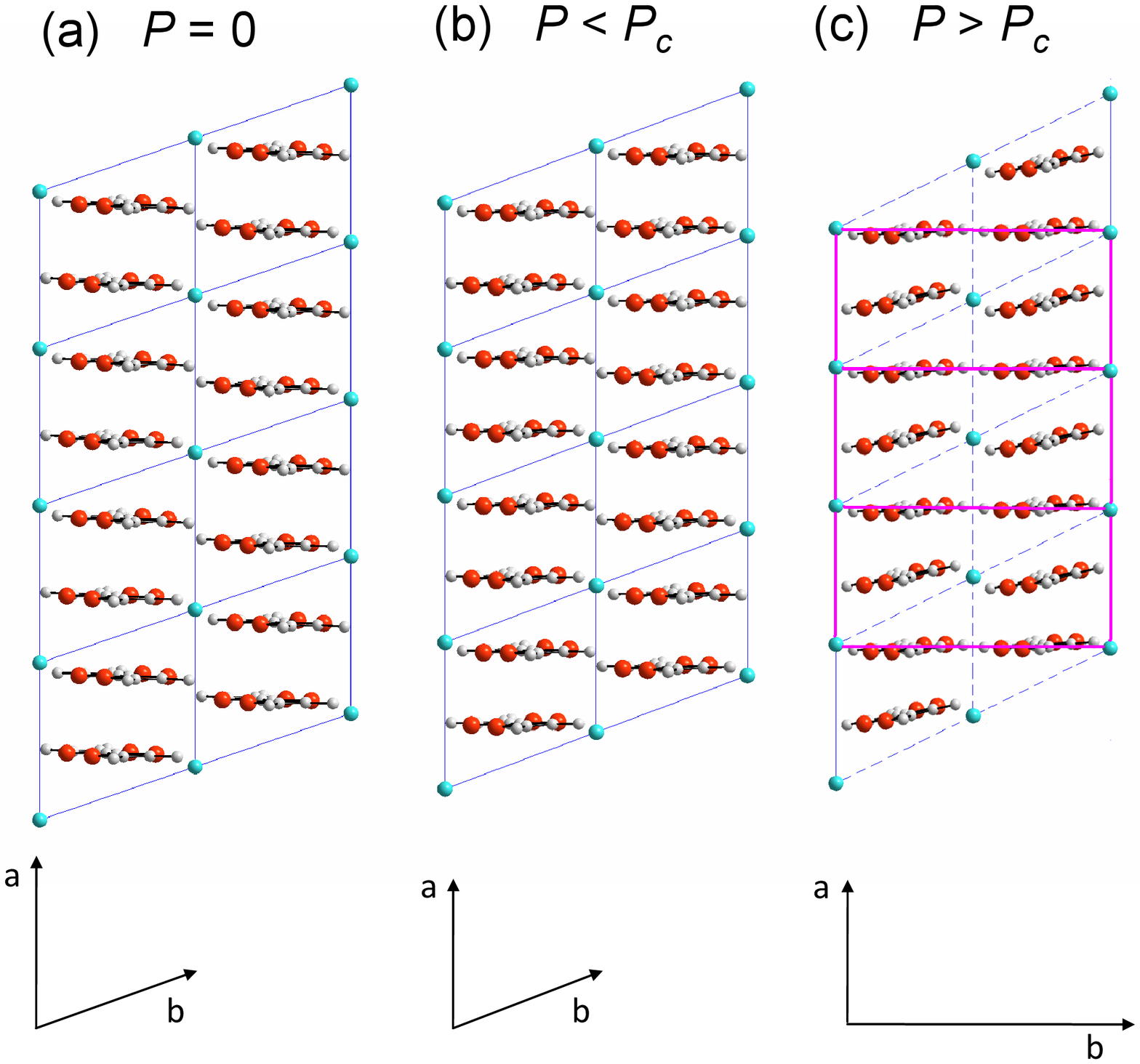}
  }\caption{(Color online) Illustration of the changes in the crystal
  structure of the Bechgaard-Fabre salts across the structural phase
  transition. (a) Crystal structure at ambient pressure; (b) compressed
  structure just below the transition pressure $P_c$; (c) new structural
  phase stabilized above the transition pressure $P_c$. Arrows depict
  translation vectors.}\label{fig:transition}
\end{figure}

We suggest a tentative model illustrated in Fig.~\ref{fig:transition}
to explain the pressure-induced structural distortion. By looking along the
$c$ axis, the figure basically shows the projection of the crystal
structure onto the $ab$ plane for pressures above and below
the transition $P_c$. The fluorine and hydrogen atoms are
not shown for simplicity. The initial compression of the crystal
lattice pushes the molecular stacks of cations closer to each other.
This effect is in particular strong along the $b$ axis, since the
interstack separation is already small at ambient pressure compared
to the separation in $c^*$ direction. Above the critical pressure
$P_c$ the interstack interaction becomes strong enough to drive the
structural instability. It leads to a small tilting of the cation
molecules in the stacks such that two neighboring unit cells become
inequivalent. As a result the unit cell doubles and the new unit
cell translational vector $b$ is defined as shown in
Fig.~\ref{fig:transition}. However, we have to stress that although
the unit cell parameters depicted in Fig.~\ref{fig:transition}
corresponds to our experimental data taken for the \SF\ sample, the
atomic positions could not be unambiguously extracted from the x-ray
diffraction data and, therefore, the depicted structural distortion
is only schematic.
\begin{figure}[t]
  \centerline{
  \includegraphics[width=0.75\columnwidth]{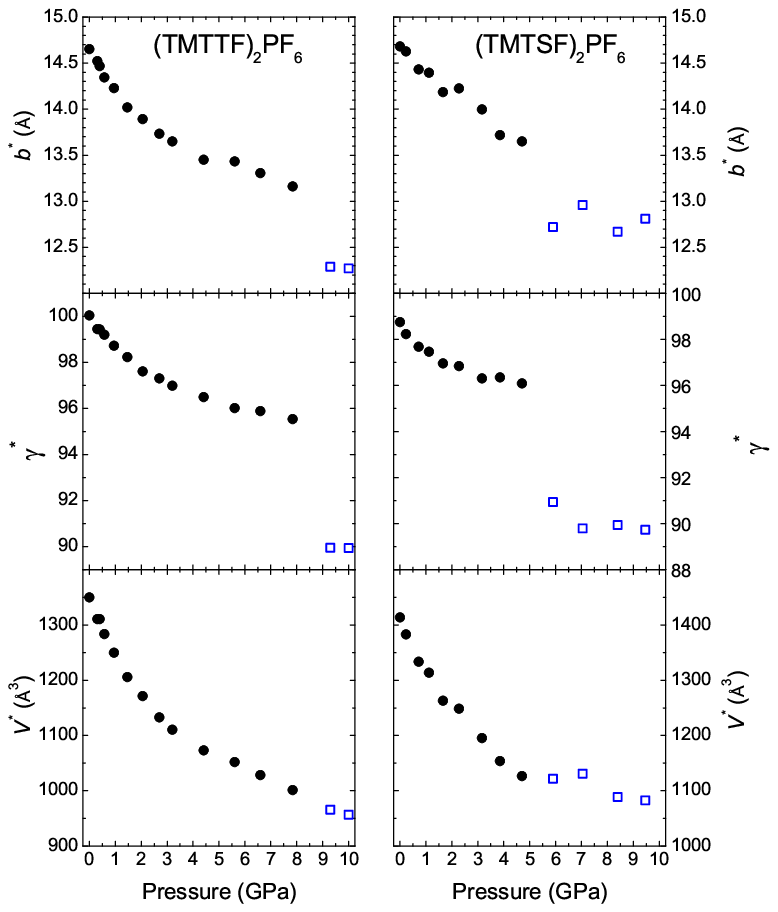}
  }\caption{(Color online) The parameters for the doubled quasi-orthorhombic unit cell of \TF\ and \SF\ as a function of
  pressure up to 10~GPa. Open symbols correspond to the parameters in the high-pressure phase directly obtained from the x-ray diffraction analysis.
   The full symbols are the parameters calculated using Eq.~\ref{eq:newparam} in the low-pressure phase.}\label{fig:newparam}
\end{figure}

The unit cell parameters of the new structure are directly related
to the unit cell parameters in the low pressure phase:
\begin{eqnarray}
 b^* &= &\sqrt{a^2+4b^2-4ab\cos\gamma} \nonumber \\
  \gamma^* &= &180-\arcsin\left(\frac{2b\sin\gamma}{b^*}\right) \\
  V^*& =& 2V \nonumber
  \label{eq:newparam}
\end{eqnarray}

Thus, in order to verify the real discontinuity, i.e., the one which
is not related to the new definition of the unit cell, we have to
compare the unit cell parameters defined as in Eq.~\ref{eq:newparam}
at pressures below and above $P_c$. Fig.~\ref{fig:newparam} shows
such a comparison of $b^*$, $\gamma^*$ and $V^*$ for both studied
salts. One can observe the clear discontinuities in $b^*$ and
$\gamma^*$ indicating that the structural phase transition is of
first order. The anomaly in the unit cell volume $V^*$
is almost invisible due to the increase in $c$ axis parameter (see
Fig.~\ref{fig:ac-transition}) which compensates decrease in $b^*$.

The reason for the unit cell doubling and formation of the
quasi-ortho\-rhombic structure could be the enhancement of the
interaction between the cation molecules in the stacks and the anions,
which becomes particularly strong for $\gamma^* = 90^\circ$, i.e.,
when the planar cation molecules and the anions arranged within the
same crystallographic plane.

Remarkably, the pressure offset between the structural transitions
in \TF\ compared to \SF\ is about 3 GPa, i.e., almost exactly the same as
the offset between electronic phases of both compounds in the
generic phase diagram of the Bechgaard-Fabre
salts.\cite{Bourbonnais98} However, it seems unlikely that the
structural distortion is related to some kind of electronic
instability, since in this case the spectrum of electronic
excitations would change. This is, however, not observed in infrared
spectra  of \SF\ above the structural transition,
i.e., for $P>5.5$~GPa. The reflectivity along the $a$ and $b'$
direction shows a metallic character with a rather small anisotropy.
No indication of an energy gap and other sudden changes was
observed \cite{Pashkin09}.

\section{Summary}

We performed room temperature x-ray diffraction study under pressure
of the quasi-one-dimensional salts \TF\ and \SF. The pressure
dependence of the unit cell constants has been obtained for
pressures up to 10~GPa. A structural phase transition from triclinic
to nearly orthorhombic phase is observed at 5.5 and 8.5~GPa in \SF\
and \TF, respectively. The transition is accompanied by a doubling
of the unit cell in direction of the $b$ axis. Several unit cell
parameters ($b^*$, $\gamma^*$ and $c$) show a considerable
discontinuity across the transition pressure indicating the phase
transition of the first order. The tentative model of the structural
distortion related to the modulation of the cation tilting in the
stacks is proposed.

\section{Acknowledgements}

We would like to thank G. Untereiner for crystal growth and N.
Drichko, M. Dumm and E. Rose for fruitful discussions and
comments. We acknowledge the ESRF facility for the provision of
beamtime. Financial support is provided by the DFG (Emmy Noether
program, SFB 484, DR228/27).

\end{document}